\documentclass[aps,prl,twocolumn,groupedaddress]{revtex4}
\usepackage{graphicx}

\begin{document}

%\preprint{}

\title{Sound propagation in a Fermi gas near a Feshbach resonance}



\author{J.~Joseph, B.~Clancy, L.~Luo, J.~Kinast, A.~Turlapov, and  J.~
E.~Thomas}
\email[jet@phy.duke.edu]{}
%\thanks{}
\affiliation{Duke University, Department of Physics, Durham, North
Carolina, 27708, USA}

\date{\today}

\begin{abstract}

Sound waves are observed and studied in an optically trapped
degenerate Fermi gas of spin-up and spin-down atoms with
magnetically tunable interactions. Measurements are made throughout
the crossover region, from a weakly-interacting Fermi gas through
the resonant Fermi superfluid regime to a Bose condensate of dimer
molecules. The measured sound velocities test the equation of state
and confirm the universal hypothesis.

\end{abstract}

%\pacs{313.43}

\maketitle

Optically-trapped, strongly-interacting Fermi
gases~\cite{OHaraScience} provide a unique laboratory for testing
nonperturbative many-body theories in a variety of fields, from
neutron stars and nuclear matter~\cite{Heiselberg} to quark-gluon
plasmas~\cite{Heinz} and high temperature
superconductors~\cite{Levin}. In contrast to these other systems,
interactions in ultra cold atomic Fermi gases are tunable using the
Feshbach resonance phenomenon~\cite{Houbiers12}: At magnetic fields
well above the resonance, one obtains a weakly attractive Fermi gas,
while well below the resonance, spin-up and spin-down atoms are
joined into dimer molecules, which form a Bose--Einstein condensate
(BEC)~\cite{GrimmBEC,JinBEC,KetterleBEC,SalomonBEC}. On resonance,
the gas is a  strongly interacting Fermi
superfluid~\cite{KetterleVortices}.

The focus of this Letter is the precision measurement of the sound
velocity at low temperatures in this unique quantum system. The
experiment addresses the rich physics of wave propagation in a
strongly-interacting system, such as possible coupling between first
and second sound~\cite{HeiselbergSound} and nonlinear behavior. We
find that the wavefront is flat despite nonuniform density across
the wavefront. This observation verifies the predicted
relation~\cite{CapuzziSFSound} between the measured velocity in a
cigar-shaped trap and the equation of state. As a result, by
measuring the velocity we probe zero-temperature equations of state
$\mu_{loc}(n)$, where $\mu_{loc}$ is the local chemical potential
and $n$ the total density of atoms. The equation of state is a
central result of nonperturbative many-body theories that determine
the thermodynamic parameters. Measurements at resonance over a wide
range of densities confirm the universal
hypothesis~\cite{OHaraScience,Heiselberg,HoUniversalThermo,ThomasUniversal}
as well as the location of the Feshbach
resonance~\cite{BartensteinFeshbach}.

The medium for the sound propagation is a 50:50 mixture of $^6$Li
atoms in the two lowest internal states or a BEC of Li$_2$ molecules
near a broad Feshbach resonance centered at $B=834$
G~\cite{BartensteinFeshbach}. The gas is cooled by forced
evaporation in an ultrastable CO$_2$ laser trap at the resonant
magnetic field~\cite{OHaraScience}. Near the end of evaporation, $B$
is adiabatically set to a chosen final value between 700 and 1100 G,
and the trap is adiabatically recompressed to a desired depth $U_0$
between 0.14 and 12 $\mu$K.

The experimental samples are nearly in the ground state, which can
be determined from the absorptive images of the
gas~\cite{OHaraScience}. The gas is imaged after release from the
trap and expanded to avoid saturation of images due to high
absorption. Close to unitarity and on the Fermi side of the Feshbach
resonance, the density profiles can be closely approximated by the
zero-temperature Thomas--Fermi profiles of a non-interacting Fermi
gas. In the molecular BEC regime, the clouds do not have any
observable thermal component, which appears when the cooling
efficiency is significantly reduced.  From the absorptive images we
measure the total number of atoms $N$ in the range $0.6$ to
$4.9\,\times 10^5$.

\begin{figure}[htb]
\includegraphics[height=3.3in]{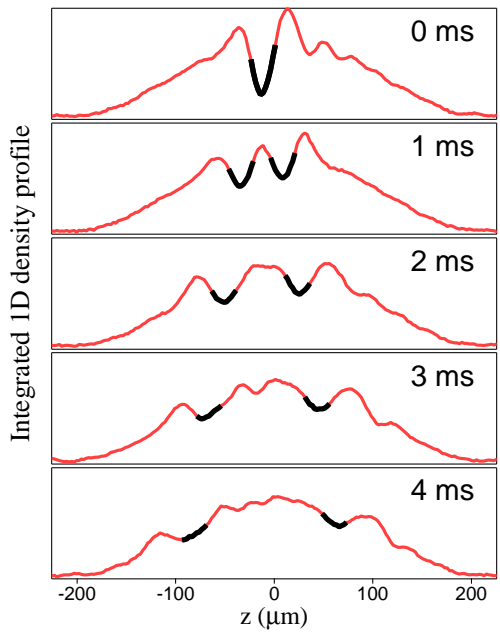}
\caption{Sound propagation at the Feshbach resonance: Integrated
axial profiles of the cloud for different in-trap propagation times.
Positions of the propagating density depressions (valleys) are
highlighted. Propagation times are shown.} \label{fig:propagation}
\end{figure}

A sound wave is excited in the sample by the same method as used for
a BEC of atoms~\cite{BoseSound}. A thin slice of green light bisects
the cigar shaped cloud. The green light at 532 nm is blue detuned
from the 671 nm transition in lithium, creating a knife which
locally repels the atoms. The $1/e$ thickness of the knife is
$\simeq 30$ $\mu$m. The height of the knife potential, $U_{knife}$,
is estimated from the size, power, and atomic polarizability.
$U_{knife}$ is chosen to be between 0.5  and 14 $\mu$K. The knife is
pulsed on for 280 $\mu$s, much shorter than typical sound
propagation times $\sim 10$ ms. The resultant perturbation
propagates outward along the axial coordinate $z$. We allow
propagation for a variable amount of time and then release the
cloud, let it expand for a fixed time, and image destructively.
Propagation of the density perturbation is shown in
Figure~\ref{fig:propagation}.  The propagating feature consists of a
region of low density (valley) with regions of high density (peak)
on either side.

\begin{figure}[htb]
\includegraphics[width=3.5in]{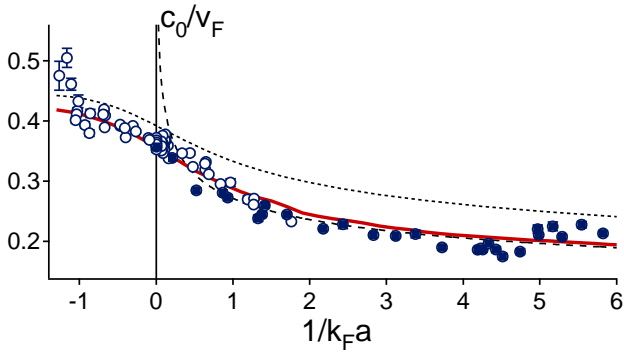}
\caption{Sound velocity in the center plane normalized to $v_F$ vs
the interaction parameter, $1/k_Fa$. Black dotted curve --
mean-field theory that uses the Leggett ground
state~\cite{AstrakharchikPrivate}. Red solid curve -- quantum Monte
Carlo calculation~\cite{AstrakharchikPrivate}. Black dashed curve --
Thomas--Fermi theory of a molecular BEC, using $a_{mol}=0.6\,a$.
Open and closed circles represent ranges of trap depths, 0.6 to 12
$\mu$K and 140 to 410 nK, respectively.}
 \label{fig:c_vs_B}
 \end{figure}

The main result of the Letter, shown in Figure~\ref{fig:c_vs_B},
is the measurement of $c_0/v_F$, the normalized sound velocity in
the center plane (at $z=0$) vs $1/k_Fa$, the dimensionless
interaction parameter.
The reference Fermi velocity $v_F$ is that of a non-interacting
Fermi gas at the trap center, i.e.,
$E_F=\hbar(\omega_x\omega_y\omega_z)^{1/3}\,(3N)^{1/3}=m_{atom}v_F^2/2=\hbar^2k_F^2/2m_{atom}$,
where $E_F$ is the Fermi energy and $k_F$ is the corresponding Fermi
wave vector. The trap frequencies $\omega_i$ are individually
calibrated for each velocity measurement. At zero temperature,
$1/k_Fa$ is the only quantity that determines the physics. The
s-wave scattering length $a=a(B)$ is estimated from
Ref.~\cite{BartensteinFeshbach}.  At $|k_Fa|\gg 1$ the system is in
the strongly interacting regime and is a superfluid. The BEC (weakly
interacting Fermi) regime corresponds to $1/k_Fa>1$ ($1/k_Fa<-1$).
We tune $k_Fa$ by changing $B$ and $U_0$.

\begin{figure}[htb]
\includegraphics[width=3.3in]{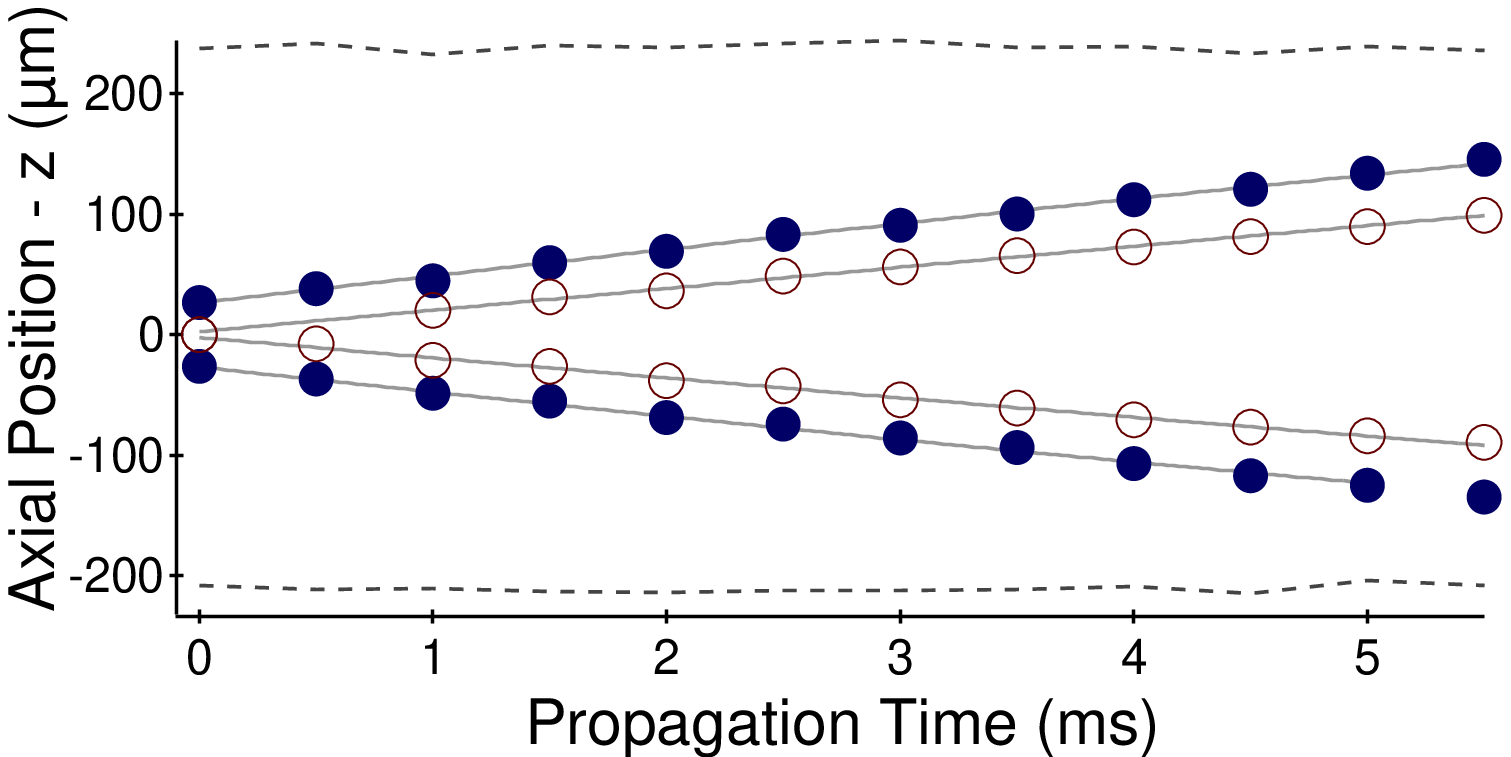}
\caption{Axial positions $z$ of the left- and right-traveling
density features of ~\ref{fig:propagation}vs the in-trap
propagation time. Coordinates of the valleys (blue-closed) and
front peaks (red-open) are shown. Dashed lines correspond to the
measured Fermi radii ($z = \pm R_z$).} \label{fig:positionvstime}
\end{figure}

Five measures are taken to reduce systematic errors. First, by
tracking both right- and left-propagating features
(Fig.~\ref{fig:positionvstime}) we eliminate effects due to the
overall motion of the cloud.
Second, we correct for a small change of the axial cloud size
between the release and imaging ($<4.7\%$, $1.2\%$ on average). In
order to obtain in-trap positions of the propagating features, each
imaged position is scaled by a coefficient of expansion calculated
within a hydrodynamic model~\cite{OHaraScience,Menotti}.

Third, we account for the dependance of the velocity on the
non-uniform density distribution along $z$. Making a general
approximation for the equation of state $\mu_{loc}(n)=C\,n^{\gamma}$
and calculating the velocity within using  hydrodynamic
theory~\cite{CapuzziSFSound,MachidaSound}, one finds that
$c(z)\propto \sqrt{1-z^2/R_z^2}$ for a harmonically trapped gas,
where $R_z$ is the axial Thomas-Fermi radius. Then, one may show
that a small density perturbation propagates as $z(t)=R_z\sin(\pm
c_0t/R_z+\varphi)$, where the phase $\varphi=\arcsin(z_{knife}/R_z)$
depends on the excitation position. $c_0$, the velocity of the
features at $z=0$, is obtained by fitting the coordinates of the
propagating features which lie within the central 65\% of the cloud
(Fig.~\ref{fig:positionvstime}).

\begin{figure}[htb]
\includegraphics[width=3.3in]{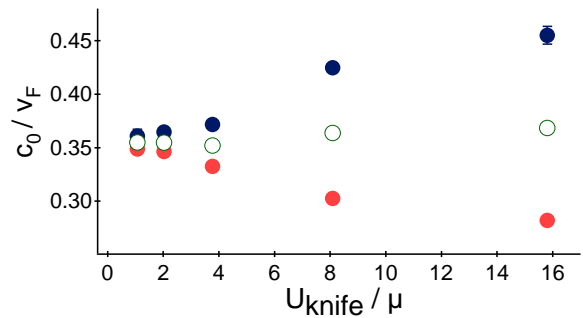}
\caption{Valley (red-closed) peak (blue-closed) and mean
(green-open) velocity vs excitation potential normalized to the
global chemical potential. Measured at the Feshbach resonance.}
\label{fig:SpeedVsPower}
\end{figure}

Fourth, we account for nonlinearity in propagation: The front peak
moves faster than the valley, as predicted earlier for
BECs~\cite{DamskiShockWavesBEC}. In Fig.~\ref{fig:SpeedVsPower}, we
show the velocities of the valley and peak vs the excitation
strength, $U_{knife}/\mu$, where $\mu$ is the global chemical
potential estimated using Monte-Carlo
modelling~\cite{Astrakharchik,AstrakharchikPrivate}.
The peak and valley velocities vary with excitation strength, but
the mean velocity remains constant.  All velocities converge to a
single value at low excitation strength showing that the mean
accurately represents the velocity of an infinitely small
perturbation. In reporting data, we use the mean value. For data of
Fig.~\ref{fig:c_vs_B}, $U_{knife}/\mu$ ranges from 1 to 30. We do
not observe formation of shock-waves which are predicted for a
BEC~\cite{DamskiShockWavesBEC}.

Fifth, the data is corrected for anharmonicity of the trapping
potential. We correct the measured sound velocities and find the
velocity values which would be measured in a perfectly harmonic
trap. Only the radial anharmonicity is taken into account because
for shallow traps, in which the correction is greatest, harmonic
magnetic confinement dominates axially. The radial confinement is
optical, and the potential is gaussian. Axially, the potential is a
sum of an optical and a parabolic magnetic potential with a common
center ($\omega_z^2=\omega_{z,\,opt}^2+\omega_{z,\,B}^2$). The
corrections for $c_0$ have been calculated to the first order in
$\mu/U_0$ from Eq.~\ref{eq:cCapuzzi} below using either a Fermi
($\gamma=2/3$) or Bose ($\gamma=1$) equation of state
$\mu_{loc}\propto n^{\gamma}$. The respective corrections are
$c_0/c_0^{meas.}=1+0.10\,\mu/U_0$ and
$c_0/c_0^{meas.}=1+0.14\,\mu/U_0$. The Fermi and Bose corrections
have been used for $1/k_Fa\leq 0$ and $1/k_Fa>1$. For $0<1/k_Fa<1$,
these two corrections were linearly combined. $\mu/U_0$ ranges from
0.025 to 0.43.

Precision characterization of the trap is important for calculating
the reference Fermi speed $v_F$. The mean radial frequency
$\omega_\perp=\sqrt{\omega_x\omega_y}$ is calculated from the radial
breathing mode frequency
$\omega_{breath}$~\cite{Kinast,KinastMagDep}, which is measured for
each point of Fig.~\ref{fig:c_vs_B} on the day of the respective
sound measurement. $\omega_{breath}$ is measured either at
resonance, where $\omega_{breath}=\sqrt{10/3}\,\omega_\perp$, or in
the far BEC regime, which yields consistent results. In the latter
case, we assume $\omega_{breath.}=2\omega_\perp$ neglecting a small
beyond-mean-field
correction~\cite{AstrakharchikFrequency,GrimmLeeHuangYang}. The
ratio $\omega_x/\omega_y=1.056$ and the total axial frequency
$\omega_z$ has been measured using parametric resonance in a
weakly-interacting Fermi gas~\cite{Kinast}. The frequency of the
magnetic field potential, $\omega_B/2\pi=20.5$ Hz at 834 G, is
measured using a sloshing mode. The trap anisotropy ratio
$\omega_\perp/\omega_z$ is between 4.1 and 25 depending on $U_0$.
For $U_0=6.1$ $\mu$K, $\omega_\perp/2\pi=690(2)$ Hz and
$\omega_z/2\pi=34.2(0.2)$ Hz. All radial frequencies have been
corrected for
anharmonicity~\cite{Stringarishift,Kinast,KinastMagDep}. The
individual corrections to $c_0$ and $v_F$ partially cancel when
calculating the ratio $c_0/v_F$. The resulting correction to
$c_0/v_F$ is within 3.1\% and, on average, is 0.7\%.

\begin{figure}[htb]
\includegraphics[width=3.3in]{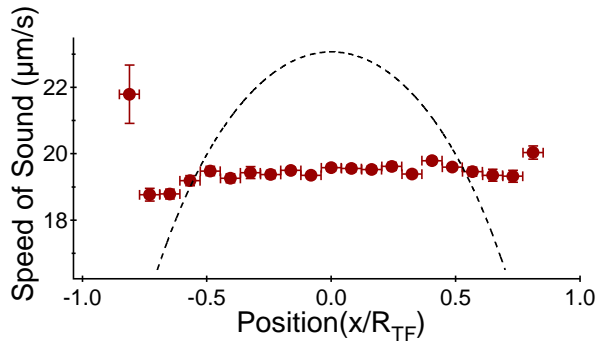}
\caption{Local sound velocity vs transverse coordinate x, in units
of the Thomas-Fermi Radius, at the feshbach resonance showing flat
wavefront. Images were divided into 21 longitudinal segments of
equal width. For each $i$th segment, $c_i(z=0)$ has been measured.
The dashed curve represents the line of site averaged isotropic
sound speed.}
 \label{fig:c_vs_x}
 \end{figure}

Linking the speed of sound to the equation of state enables
quantitative tests of theories. By analyzing the wave front shape we
now show that the hydrodynamic model of Capuzzi et
al.~\cite{CapuzziSFSound}, for a cigar-shaped trap, provides a
correct description of propagation. The primary result of this model
is that the wave front stays flat, as described by
\begin{equation}
c(z)=\left(\frac{1}{m_{atom}}\frac{\int n\,dx\,dy}{\int
\left(\partial\mu_{loc}/\partial
n\right)^{-1}\,dx\,dy}\right)^{1/2}. \label{eq:cCapuzzi}
\end{equation}
In contrast, a simple model of isotropic sound propagation assumes
$c_{loc}(n)=\sqrt{(\partial P/\partial n)/m_{atom}}$, where $P$ is
the pressure. In this case, the wave front curves as the wave
propagates. For the same equation of state, these two theories
predict a few per cent difference in the speed of the
perturbation. In Figure~\ref{fig:c_vs_x} we experimentally
demonstrate that the speed is independent of the transverse
position, which validates Eq.~\ref{eq:cCapuzzi} for our
experimental conditions. And, hence, we can rely on
Eq.~\ref{eq:cCapuzzi} in connecting $c_0$ to $\mu_{loc}(n)$.
At this point we are in a position to compare the data of
Fig.~\ref{fig:c_vs_B} to predictions.

In the BEC regime, at $1/k_Fa>1$, the experiment tests the
interaction properties of composite bosons, molecules made out of
two fermions. One may approximate the state of the gas as a BEC of
singlet molecules that collide with s-wave scattering length
$a_{mol}$. Using the local equation of state
$\mu_{loc}(n_{mol})=2\pi\hbar^2a_{mol}n/m_{mol}$ in
Eq.~\ref{eq:cCapuzzi}, one finds
$c_0/v_F=(5/2\,\,k_Fa_{mol})^{1/5}/4$. An extension of the Leggett
ground state into the Bose regime~\cite{NozieresMFTheory} gives
$a_{mol}=2\,a(B)$ resulting in a prediction for the speed $25-40\%$
higher than measured. An exact solution of the four-fermion
scattering problem predicts $a_{mol}=0.6\,a$~\cite{Petrov} yielding
a speed (black dashed curve in Figure~\ref{fig:c_vs_B}) close to the
measurements.

Theories covering the whole crossover from the Fermi to Bose regime
can be tested against our data. In Fig.~\ref{fig:c_vs_B}, the black
dotted curve~\cite{AstrakharchikPrivate} represents the theory based
on the Leggett ground state~\cite{NozieresMFTheory}. The theory
predicts significantly higher speed than measured except for the
Fermi side of the resonance. The red solid
curve~\cite{AstrakharchikPrivate} shows the calculation that uses
the equation of state obtained from a Monte--Carlo
simulation~\cite{Astrakharchik}. This theory shows much better
agreement with the data. Interestingly, in most of the molecular BEC
regime, at $1/k_Fa>1$, the data is systematically lower than the
prediction.

The lower measured speed at $1<1/k_Fa<5$ could be due to the
coupling of first and second sound, which travels
slower~\cite{HeiselbergSound}. High temperature cannot be a reason
for lower speed. We have found that higher temperatures correspond
to higher sound velocities. It is also unlikely that the speed is
lower due to a dark soliton, which travel slower than sound, and, in
principle may exist in our molecular BECs. The tracked density
depressions are most likely not dark solitons for three reasons.
First, unlike a dark soliton, the feature spreads out and loses
contrast with time (Fig.~\ref{fig:propagation}), which is observed
at all $1/k_Fa$'s. Second, the soliton size is predicted to be of
the same order as the coherence
length~\cite{JacksonDarkSolitonTeory},
$\hbar/\sqrt{2\mu\,m_{mol}}<1$ $\mu$m, while the imprinted density
depression has the width of $10-20$ $\mu$m. Third, the optimum
soliton excitation requires phase imprint of
$\pi$~\cite{BurgerDarkSolitonExp}, while in our experiment, the
phase imprint for most of the BEC-side experiments is $\simeq
20\pi$.

The small speed increase at $1/k_Fa>5$ could occur due to a
technical reason: To achieve these high $1/k_Fa$ values we have to
use very shallow traps. The BECs released from these weakest traps
are small and difficult to image.

The universal hypothesis is positively tested in our measurements.
The hypothesis states that at resonance, where $k_Fa=\infty$,
$c_0/v_F$ should be independent of density. Indeed, within the
experimental precision, at resonance, at 834(2)G, we observe that
$c_0/v_F$ remains constant as the trap depth is changed from 410 nK
to 12 $\mu$K. In contrast, below the Feshbach resonance, at 821 G,
$c_0/v_F$ decreases with decreasing $U_0$. These measurements of
$c_0/v_F$ vs $U_0$ show that the Feshbach resonance location is much
closer to 834~G than to 822~G~\cite{SchunckFeshbach}.

From the sound velocity at resonance, we have also measured the
universal constant $\beta$~\cite{Heiselberg,OHaraScience}, which
connects the local interaction and Fermi energy in the universal
regime: $U_{int}=\beta\,\epsilon_{F}$. Using
$\mu_{loc}=\epsilon_F+U_{int}=(1+\beta)\,(3\pi^2n)^{3/2}\,\hbar^2/2m_{atom}$
and Eq.~\ref{eq:cCapuzzi}, one may relate $c_0$ to $\beta$ as:
\begin{equation}
\frac{c_0}{v_F}=\frac{(1+\beta)^{1/4}}{\sqrt{5}},
\label{eq:soundres}
\end{equation}
from which we find $\beta=-0.570(0.015)$.

On the far Fermi side of the Feshbach resonance, at $1/k_Fa<-1.3$,
propagation of sound is not observed. Instead, the hole from the
laser knife fills without propagating, as in a non-interacting gas.
At the temperatures we achieve, the gas is not necessarily a
superfluid in the weakly interacting Fermi gas region, and
hydrodynamic sound propagation is provided by collisional, rather
than superfluid hydrodynamics.

In conclusion, we have observed sound propagation, connected
observations to a hydrodynamic model, and provided data for
testing theories describing strongly-interacting Fermi systems as
well as their crossover into the weakly interacting Fermi and Bose
regimes.

We thank G.~Astrakharchik, P.~Capuzzi, L.~Carr, K.~Levin, Qijin
Chen, Yan He, K. Machida, and T.~Ghosh for stimulating discussions
and correspondence. This research is supported by the Physics
Divisions of the Army Research Office and the National Science
Foundation, the Physics for Exploration program of the National
Aeronautics and Space Administration, and the Chemical Sciences,
Geosciences and Biosciences Division of the Office of Basic Energy
Sciences, Office of Science, U. S. Department of Energy.

%\bibliography{FermiGas}

\begin{thebibliography}{31}
\bibitem[{Str()}]{Stringarishift}
\bibinfo{note}{These frequency shifts were derived by S. Stringari, private
  communication. In the molecular BEC regime,
  $\omega_{breath.}/2\omega_\perp=1-7/6\,m\omega_\perp^2x_{rms}^2/U_0$}.
\end{thebibliography}

\end{document}